\documentclass{elsart}
\usepackage{graphicx,amssymb}
\begin{document}
\newcommand{\be}{\begin{equation}}
\newcommand{\ee}{\end{equation}}
\newcommand{\bq}{\begin{eqnarray}}
\newcommand{\eq}{\end{eqnarray}}
\newcommand{\bsq}{\begin{subequations}}
\newcommand{\esq}{\end{subequations}}
\newcommand{\bc}{\begin{center}}
\newcommand{\ec}{\end{center}}
\newcommand {\R}{{\mathcal R}}
\newcommand{\al}{\alpha}
\newcommand\lsim{\mathrel{\rlap{\lower4pt\hbox{\hskip1pt$\sim$}}
    \raise1pt\hbox{$<$}}}
\newcommand\gsim{\mathrel{\rlap{\lower4pt\hbox{\hskip1pt$\sim$}}
    \raise1pt\hbox{$>$}}}

\begin{frontmatter}
\title{Understanding Domain Wall Network Evolution}
\author[cfp,df]{P.P. Avelino\corauthref{cor}},
\corauth[cor]{Corresponding author.}
\ead{ppavelin@fc.up.pt}
\author[cfp,df]{J.C.R.E. Oliveira},
\ead{jeolivei@fc.up.pt}
\author[cfp,cms]{C.J.A.P. Martins}
\ead{C.J.A.P.Martins@damtp.cam.ac.uk}
\address[cfp]{Centro de F\'{\i}sica do Porto, Rua do Campo Alegre 687, 4169-007 Porto, 
Portugal}
\address[df]{Departamento de F\'{\i}sica da Faculdade de Ci\^encias
da Universidade do Porto, Rua do Campo Alegre 687, 4169-007 Porto, 
Portugal}
\address[cms]{Department of Applied Mathematics and Theoretical 
Physics, CMS,\\ University of Cambridge,
Wilberforce Road, Cambridge CB3 0WA, U.K.}
\begin{abstract}
We study the cosmological evolution of domain wall networks in two and three 
spatial dimensions in the radiation and matter eras using a 
large number of high-resolution field theory simulations with a large 
dynamical range. We investigate the dependence of the uncertainty in key 
parameters characterising the evolution of the network on the size, 
dynamical range and number of spatial dimensions of the simulations 
and show that the analytic prediction compares well with the simulation 
results. We find that there is ample evidence from the simulations of a 
slow approach of domain wall networks towards a linear scaling solution. 
However, while at early times the uncertainty in the value of the scaling 
exponent is small enough for deviations from the scaling 
solution to be measured, at late times the error bars are much larger 
and no strong deviations from the scaling solution are found. 
\end{abstract}
\end{frontmatter}

\section{Introduction}
\label{intr}

Topological defects are generic in nature and may be formed whenever a 
phase transition occurs. In an expanding universe cooling down from a 
very hot initial state it is to be expected that topological defects 
may provide a unique window onto the physics of the early universe offering 
perhaps the best hope of a clear observable link between cosmology and 
particle physics \cite{Kibble,Book}.
Most cosmological studies of cosmic defects have focused on cosmic strings 
due to their interesting properties and strong motivation from fundamental 
physics (see for example \cite{Sarangi,Kibble1,Polchinski} and references 
therein). Although standard cosmic strings are now ruled out as the sole 
contribution to the large scale structure of the 
universe \cite{Bouchet,Durrer} they may still be the dominant source of 
perturbations on 
small cosmological scales and may give rise to a number of interesting 
cosmological consequences
\cite{Laix,Liddle,Barbosa}. However, at present there are only two 
observations for which cosmic strings seem to offer the most natural 
explanation \cite{Sazhin,Khovanskaya}.

Domain wall scenarios have attracted less attention 
since heavy domain walls in a linear scaling regime rapidly dominate 
the energy density of the universe. 
Moreover, domain walls which are light enough 
to satisfy current CMB constraints have a negligible direct contribution 
to structure 
formation. However, in this case a number 
of interesting consequences are possible such as a contribution to the dark 
energy \cite{Bucher,Conversi} (if domain walls are frozen in comoving 
coordinates) and a small but measurable contribution to the CMB 
anisotropies at large angular scales \cite{Friedland}. Domain walls may 
also separate regions in the universe with different values of the 
cosmological parameters and/or fundamental constants of nature 
\cite{Inhomog,Menezes}. 

In this paper we perform a quantitative study of the cosmological evolution of 
domain wall networks. We investigate the dependence of the uncertainty in key 
parameters characterising the evolution of the network on the size, 
dynamical range and number of spatial dimensions of the simulations using 
a simple analytic model.
We then compare our analytic predictions with the results of a large set of 
high-resolution simulations of domain walls in two and three spatial 
dimensions \cite{Oliveira}, using the standard Press-Ryden-Spergel (PRS) 
algorithm \cite{Press} (see also \cite{Coulson,Larsson,Fossils,Garagounis}), 
and discuss the evidence from the simulations of a slow approach towards a 
linear scaling regime. Previous studies of domain wall network evolution 
\cite{Press,Coulson,Larsson,Fossils,Garagounis} 
having a smaller number of simulations with smaller size and dynamical range
than the present one have found some hints for deviations from a
scale-invariant evolution. It is therefore crucial to investigate if
these are only transitory or if there is a more fundamental reason for such
deviations.

The present article is a follow-up of \cite{Oliveira}. There, we concentrated 
on the overall (global) dynamical features of the simulations. On the other
hand, having a large dynamic range means that a more localised analysis is
also possible, and in particular local exponents can be calculated with
relatively small errors. In the present paper we explore this possiblity,
and also make use of the large number of simulations to discuss some
analytic ways to estimate statistical errors.

\section{Domain Wall Network Evolution}
\label{dwe}

We study the evolution of a domain wall network in a flat
homogeneous and isotropic Friedmann-Robertson-Walker (FRW) 
universe. We consider a scalar field $\phi$ with the Lagrangian
density
\begin{equation}
\mathcal{L}={\frac{1}{2}}\phi_{,\alpha}\phi^{,\alpha}-V(\phi)\,,
\label{action1}
\end{equation}
and we will take $V(\phi)$ to be the generic $\phi^{4}$ potential
with two degenerate minima given by 
\begin{equation}
V(\phi)=V_{0}\left({\frac{\phi^{2}}{\phi_{0}^{2}}}-1\right)^{2}\,,
\label{potential}
\end{equation}
which obviously admits domain wall solutions.
Following the procedure described in ref. \cite{Press} we modified 
the equations of motion in such a way that the co-moving thickness 
of the domain walls is fixed in co-moving coordinates allowing us 
to resolve the domain walls throughout the full dynamical range 
of the simulations. With this modification implemented the equations 
of motion for the field $\phi$ become: 
\begin{equation}
{\frac{{\partial^{2}\phi}}{\partial\eta^{2}}}+\alpha\left(\frac{d\ln 
a}{d\ln\eta}\right){\frac{{\partial\phi}}{\partial\eta}}-{\nabla}^{2}\phi=
-a^{\beta}{\frac{{\partial 
V}}{\partial\phi}}\,.\label{dynamics2}
\end{equation}
where $a$ is the scale factor, $\eta$ is the conformal time and $\alpha$ 
and $\beta$ are constants. We take $\beta=0$ in 
order to have constant co-moving thickness and $\alpha=3$ to ensure that 
the momentum conservation law of the wall evolution in an expanding universe 
is maintained \cite{Press}. Equation (\ref{dynamics2}) is then 
integrated using a standard finite-difference scheme.

We have verified that the PRS alghoritm gives the correct results in some
special cases such as the dynamics of a plane wall or the collapse of a
spherical or cilindrical domain wall. We have also verified that it
appears to have a small impact on the large-scale dynamics of domain wall
networks and does not seem to affect the quantities we want to measure
for the purpose of testing scaling properties provided a minimum acceptable
tickness is used. However, it is only possible to test the performance of the
PRS alghoritm over a narrow window since the `true' equation of motions
for the domain walls rapidly make the wall thickness smaller than the grid
size.

In addition to these simple tests with domain walls,
the PRS algorithm has been much more extensively used and tested in the
case of cosmic strings (see for example ref. \cite{Moore}). From
those one can infer that with the use of the PRS algorithm some
quantitative features of the networks will indeed differ (for example, the
distribution of small-scale features on the networks), but the broad
features will be largely unchanged. An example of the latter is the
existence of an attractor scaling solution, and how fast it is reached
starting from some given initial configuration. Since this is the issue we
are studying here, we believe that the PRS algorithm is adequate for our
purposes.

The ratio between the kinetic and potential energy of the 
domain walls is approximately given by 
\begin{equation}
F\equiv\frac{1}{A}\sum_{i,j,k}\left(\frac{\partial\phi_{ijk}}
{\partial\eta}\right)^{2}\,.
\label{energiacinetica}
\end{equation}
where $A$ is the total co-moving area of the domain walls determined 
using the algorithm described in ref. \cite{Press,Fossils,Oliveira} and 
we are measuring length in units of the grid spacing $\Delta x$ (so that 
$\Delta x=1$). 
This quantity is related to the root-mean squared 
velocity of the domain walls which should be conserved in a linear 
scaling regime. We assume the initial value of $\phi$ to be a 
random variable between $-\phi_{0}$ and $+\phi_{0}$ and the 
initial value of $\partial\phi/\partial\eta$ to be zero. 
See \cite{Fossils,Oliveira,Moore} for further discussion of these and 
other issues.

\section{Analytic Modelling}
\label{model}

We consider a simple model for the evolution of the uncertainty in 
key parameters characterising the evolution of a domain wall network 
which we then test against domain 
wall network simulations. Let us define the comoving correlation length of the 
network as $\xi \equiv V/A$. Consider a cubic grid in $N_D$ 
dimensions where each cube of comoving volume $V_\xi=\xi^{N_D}$ has 
$N_D$ faces of its own. Note that the number of faces of a cube in 
$N_D$ dimensions is $2N_D$ but each one of them belongs to two adjacent 
cubes so that in practise each cube has $N_D$ faces of its own. 
We also assume that there is on 
average one domain wall of comoving area $\xi^{N_D-1}$ per cube of 
comoving volume $V_\xi$ 
occupying 
one of the faces of the cube. Hence, the probability that a face of such 
a cube is occupied by a domain wall is $p=1/N_D$ so that the variance 
in the number of domain walls per cube is $\sigma_\xi^2=N_D p q$ where $q=1-p$.
Hence $\sigma_\xi={\sqrt {1-1/N_D}}$. If we now consider a cube 
of comoving volume $V=L^{N_D}$ it will contain $N=V/V_\xi$ cubes of 
comoving volume $V_\xi$.
The standard deviation, $\sigma_X$, of $X=n/{\bar n}$ (where $n$ denotes  
the number of domain walls of comoving area $\xi^{N_D-1}$ and ${\bar n}$ 
its average value) on a given volume, $V$, 
is proportional to ${\sqrt N}/N$ and consequently
\begin{equation}
\sigma_X=\sigma_\xi N^{-1/2}={\sqrt {1-\frac{1}{N_D}}} 
\left(\frac{A \eta}{V}\right)^{-N_D/2} \left( \frac{\eta}{L} \right)^{N_D/2}.
\end{equation}
This can also be calculated for any other variable, $Y$, characterising the 
network if $Y$ is proportional to $X$. In this case 
\begin{equation}
\sigma_Y=\sigma_X {\bar Y}.
\end{equation}
Here the random variable $Y$ may represent the total area $A$, or the ratio 
between the kinetic and potential energy parametrised by $F$. Although in 
the first case it is a fair assumption to draw a direct relation between 
the number of domain walls and the total area, in the second case we expect 
the relation to be less direct due to the dispersion in the domain wall 
velocities and to the existence of an important fraction of the kinetic 
energy which is not associated with the domain walls themselves but with their 
decay products.

The scaling exponent may be calculated from the value of $R \equiv \eta A/V$ 
at two different values of the conformal time $\eta_1$ and $\eta_2$ as 
\begin{equation}
\lambda(\eta_1,\eta_2) \equiv \frac{\ln\left({R_1}/{R_2}\right)}
{\ln\left({\eta_1}/{\eta_2}\right)} \sim 
\frac{\ln\left({{\bar R}_1}/{{\bar R}_2}\right)}
{\ln\left({\eta_1}/{\eta_2}\right)} + 
\frac{\Delta_1-\Delta_2}{\ln\left({\eta_1}/{\eta_2}\right)}
\end{equation}
so that
\begin{equation}
\sigma_\lambda \sim \frac{{\sqrt {\sigma^2_{\Delta_1}+\sigma^2_{\Delta_2}}}}{\ln\left({\eta_1}/{\eta_2}\right)}.
\label{slambda}
\end{equation}
assuming that $\Delta_1$ and $\Delta_2$ are small (compared 
with ${\bar R}_1$ and ${\bar R}_2$ respectivelly) and uncorrelated.
Here $\Delta=(R-{\bar R})/{\bar R}$ where ${\bar R}(\eta)$ 
denotes the average value 
of $R(\eta)$ over a given number, $N_S$, of simulations. 
We also may also estimate $\sigma_Y$ for several key parameters, 
$Y=R,F,\lambda$, describing the evolution of the network directly 
from the simulations as
\begin{equation}
s_Y^2=\frac{1}{N_S-1}\sum_{i=1}^{N_S} (Y_i- {\bar Y})^2.
\end{equation}
Note that the average value of $Y$ for a given sample of simulations, 
${\bar Y}$, is the best estimator of the real average of $Y$ with and 
error of $\sigma_Y/{\sqrt N_S}$. Also $s_Y$ is the best estimator of 
$\sigma_Y$ and $s_Y \sim \sigma_Y$ if $N_S$ is large enough.

\section{Results and discussion}
\label{resdis}

\begin{figure}
\includegraphics[width=2.8in]{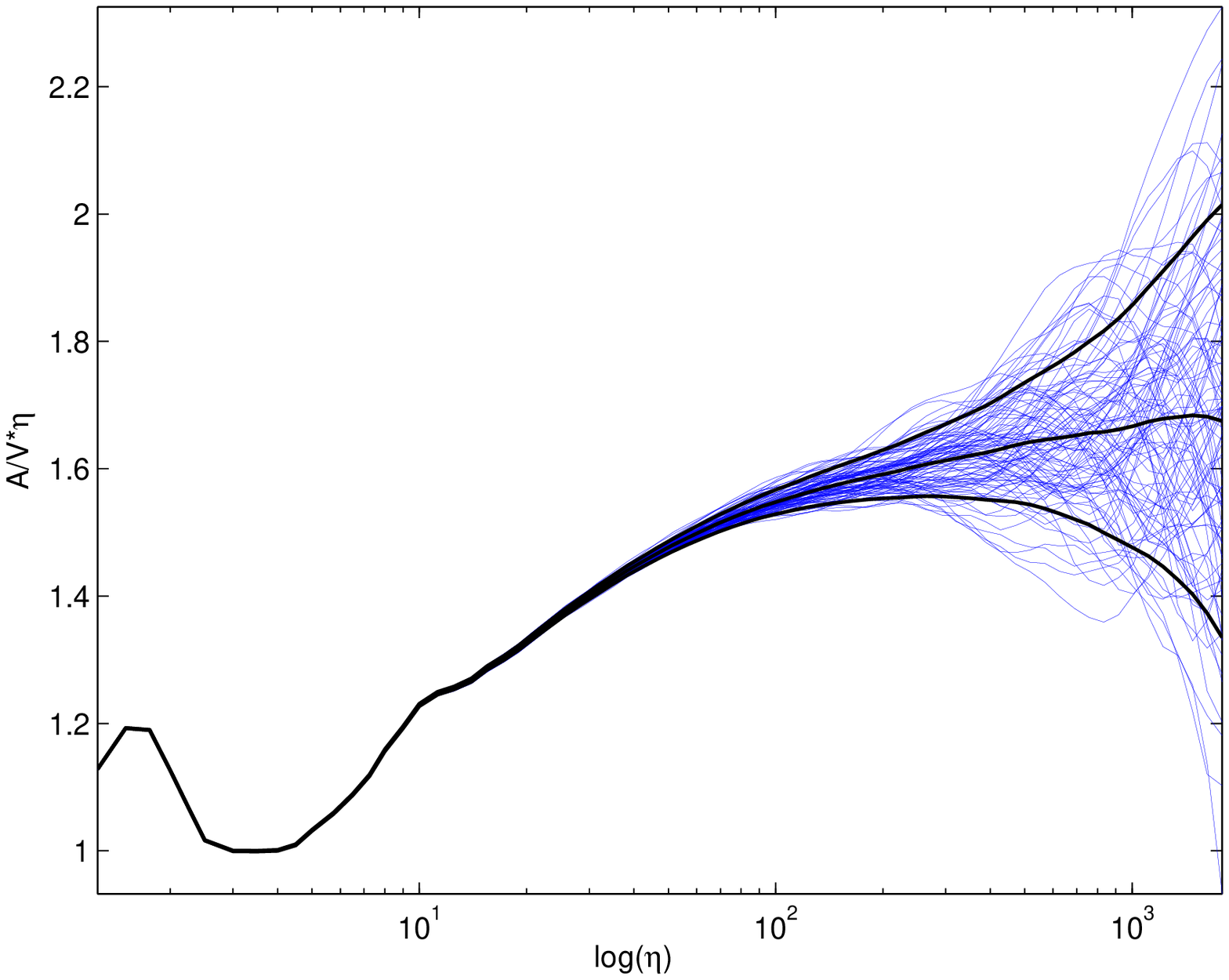}
\includegraphics[width=2.8in]{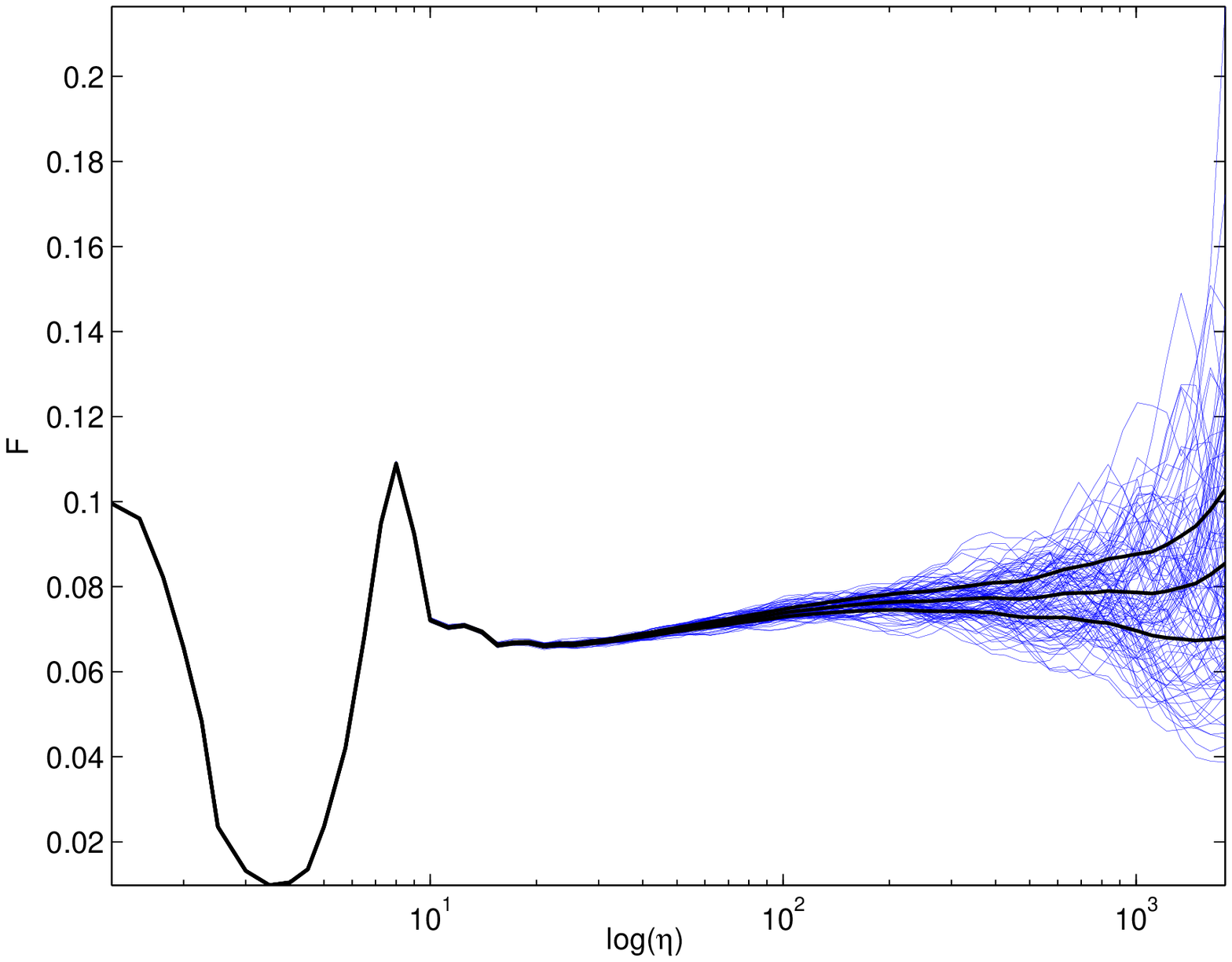}
\caption{\label{fig1}  The evolution of $R \equiv \eta A/V$ and $F$ for 
one hundred $4096^2$ $2D$ matter era simulations. We also plot the average 
evolution of these parameters and the expected 1-sigma interval around the 
mean (using the analytic estimates for $\sigma_R$ and $\sigma_F$ 
described in Sec. III).}
\end{figure}

\begin{figure}
\includegraphics[width=2.8in]{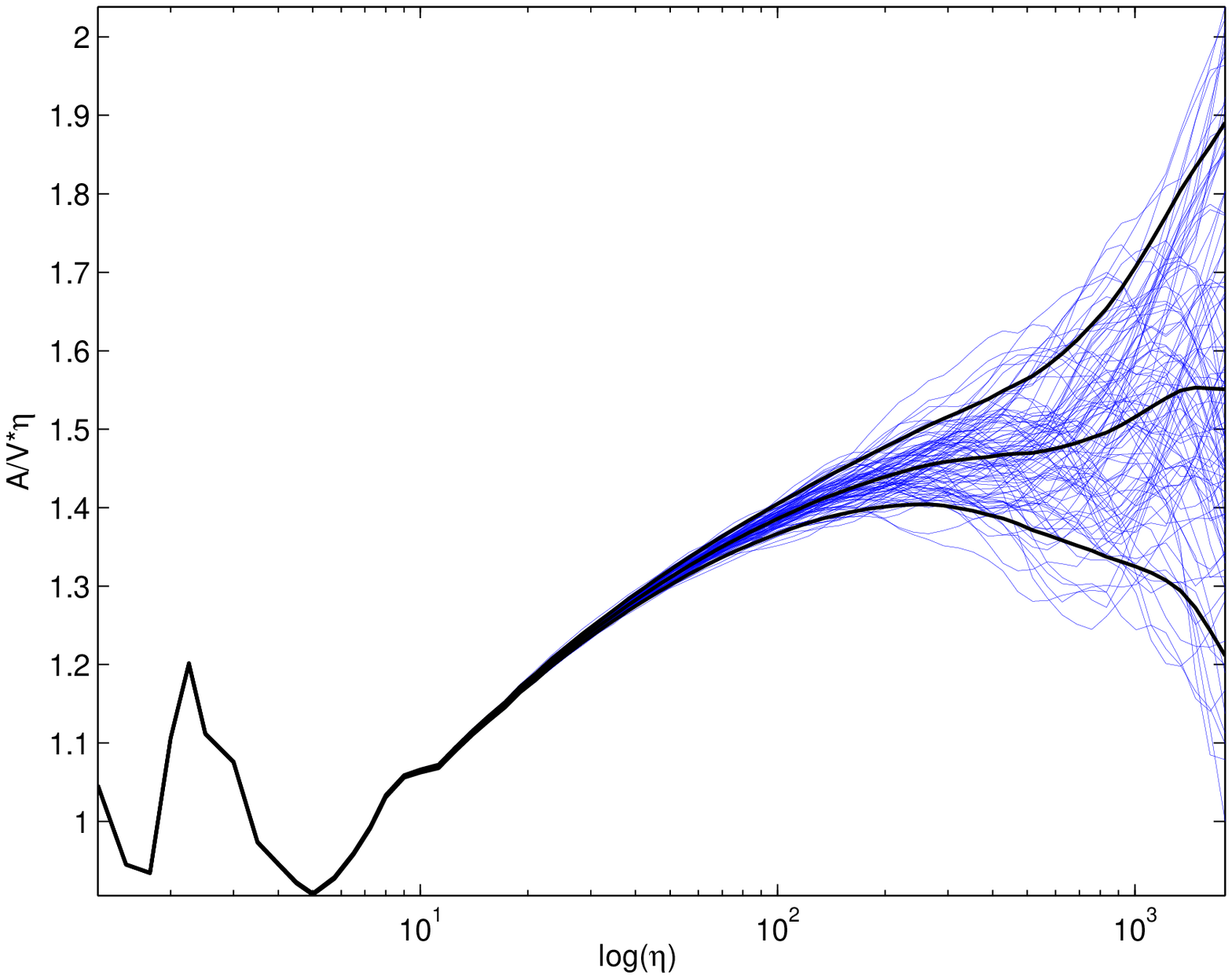}
\includegraphics[width=2.8in]{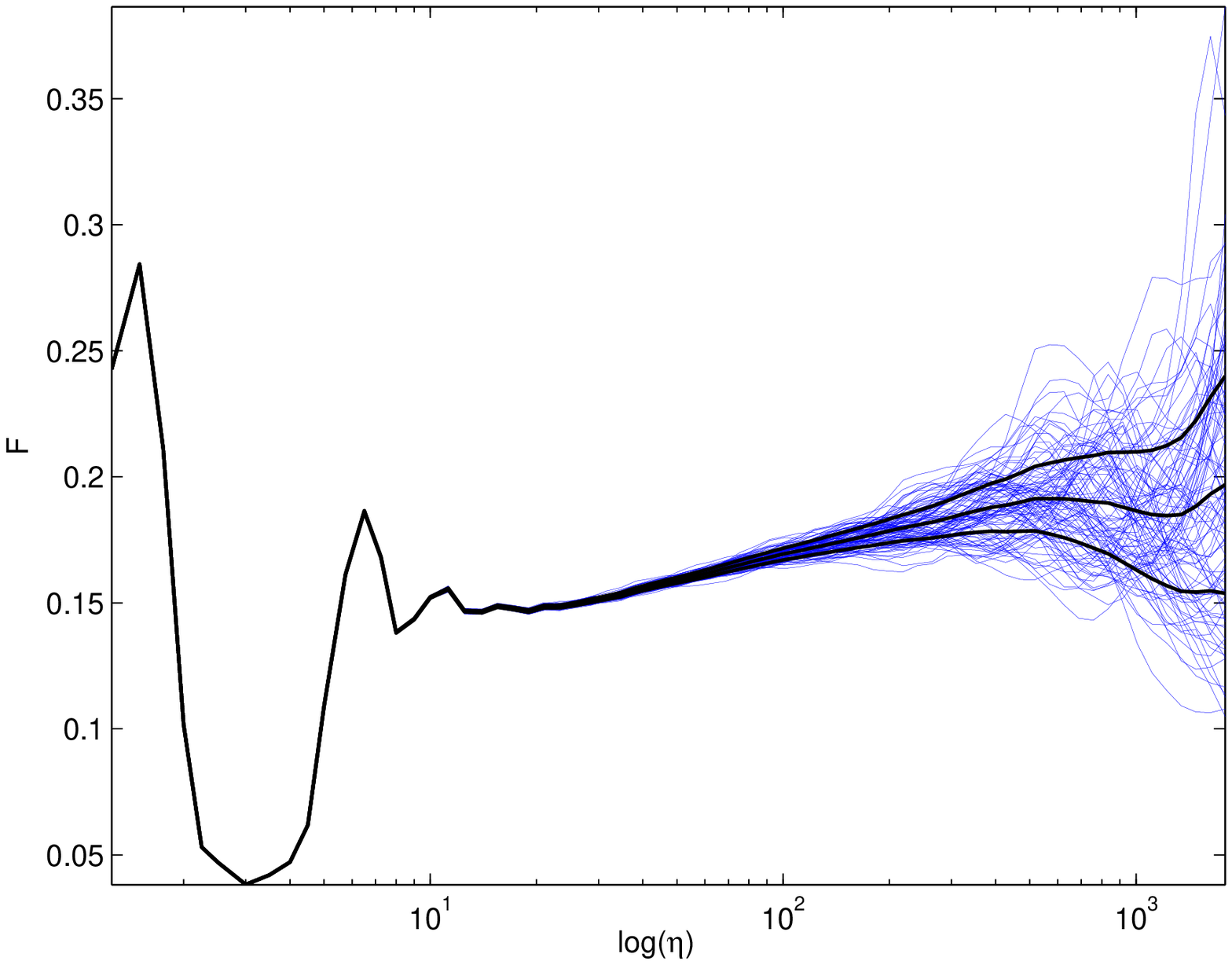}
\caption{\label{fig2} The same as Fig. \ref{fig1} but now for one hundred  
$4096^2$ $2D$ radiation era simulations.}
\end{figure}

\begin{figure}
\includegraphics[width=2.8in]{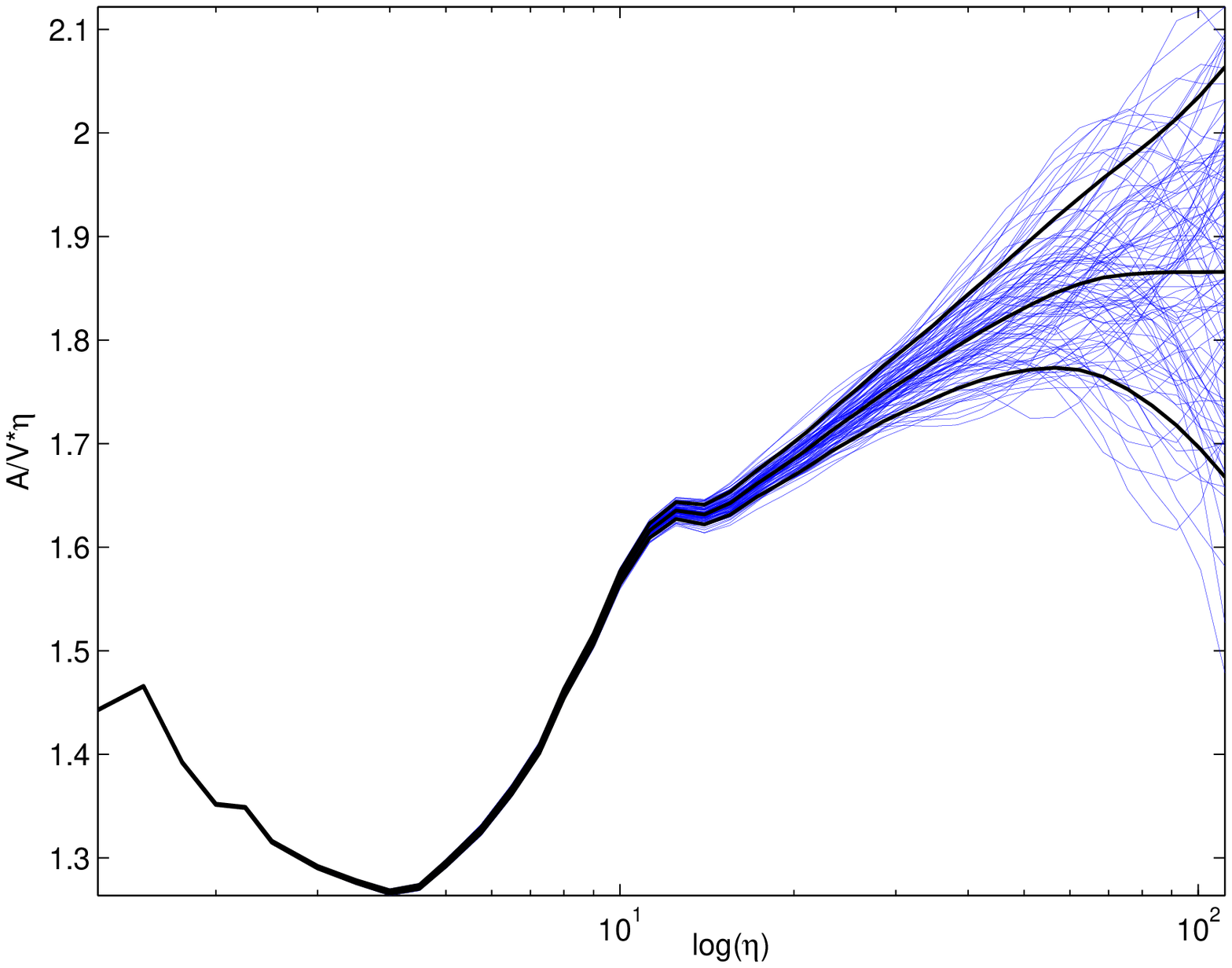}
\includegraphics[width=2.8in]{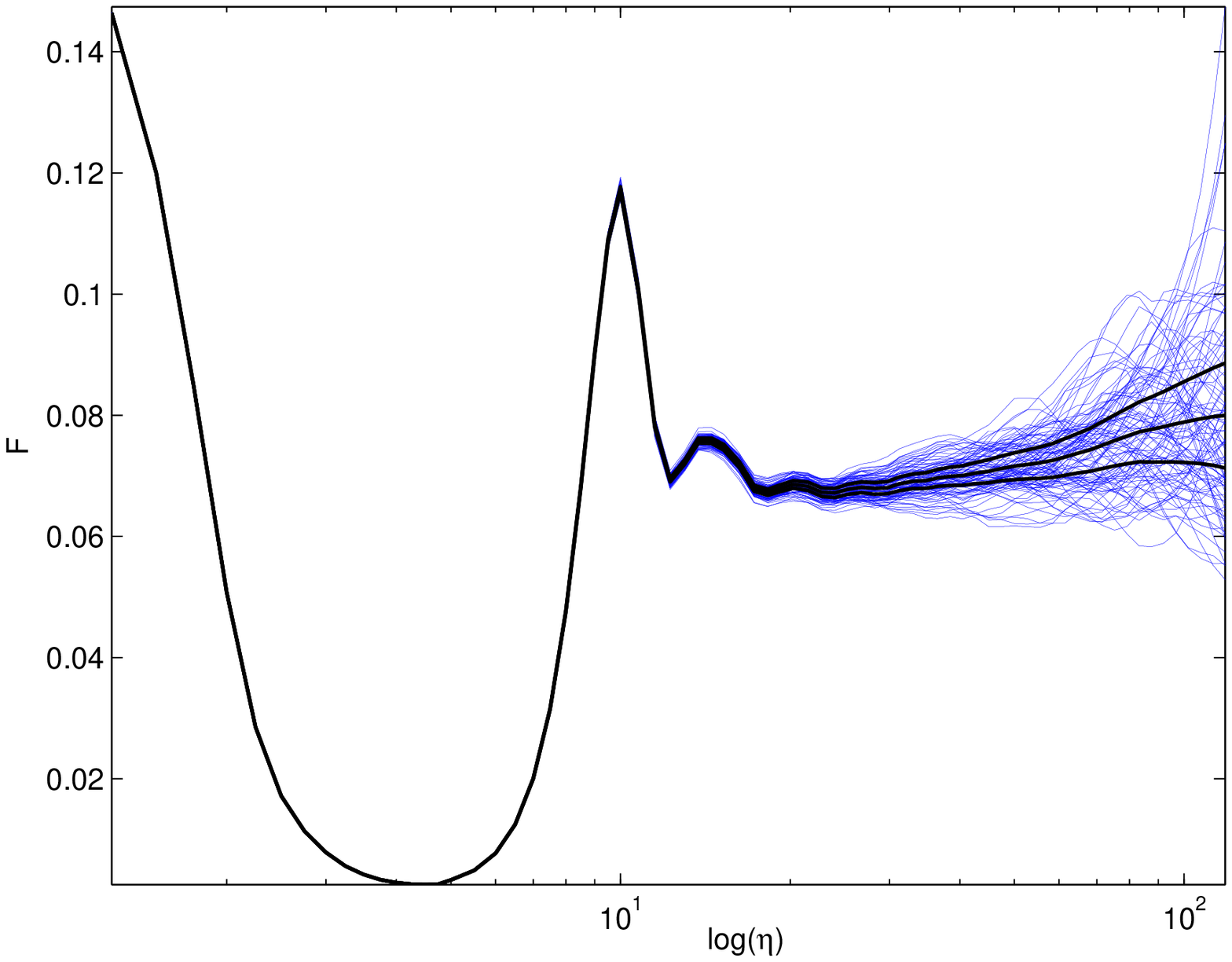}
\caption{\label{fig3}  The same as Fig. \ref{fig1} but now for one hundred 
$256^3$ $3D$ matter era simulations.}
\end{figure}

\begin{figure}
\includegraphics[width=2.8in]{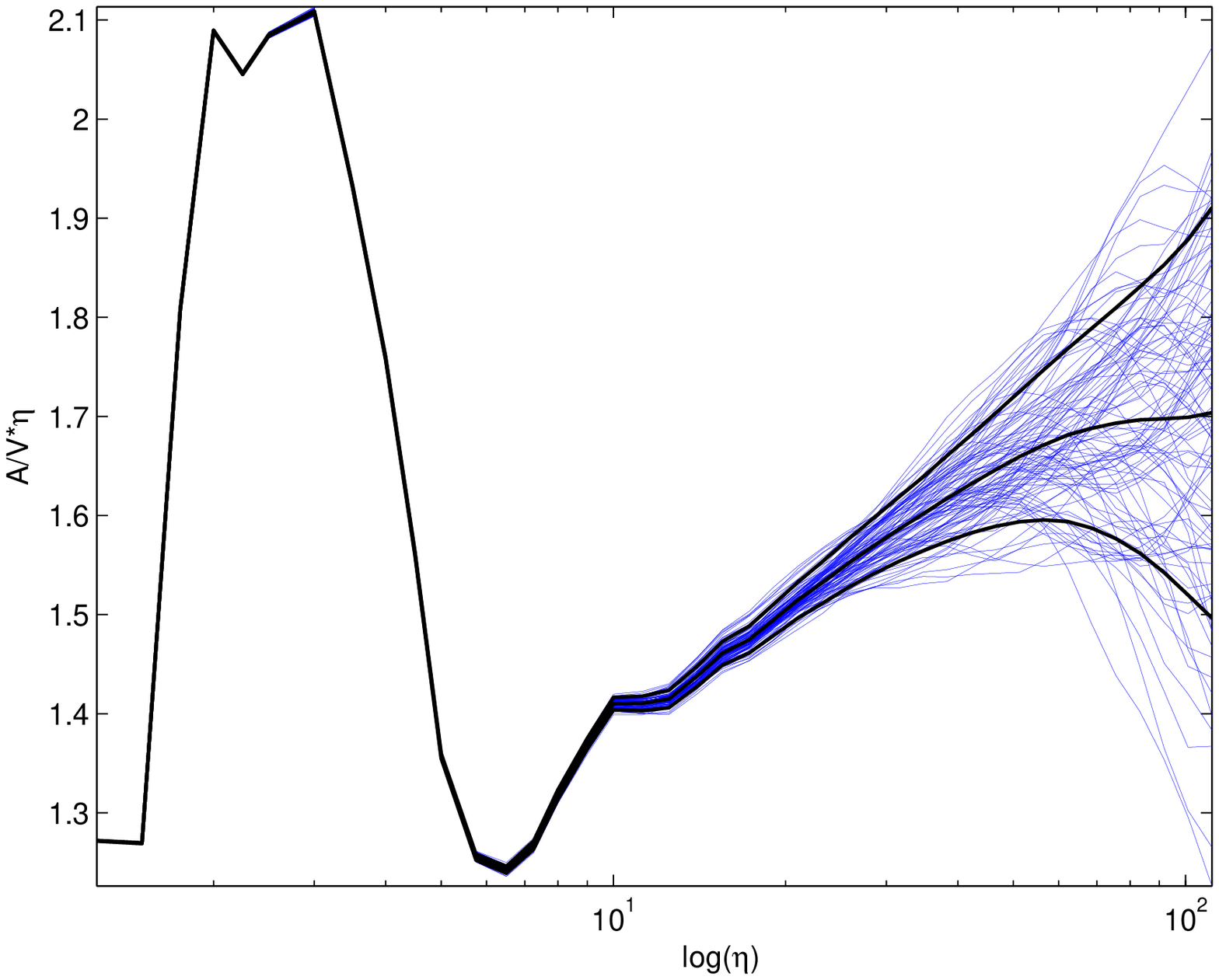}
\includegraphics[width=2.8in]{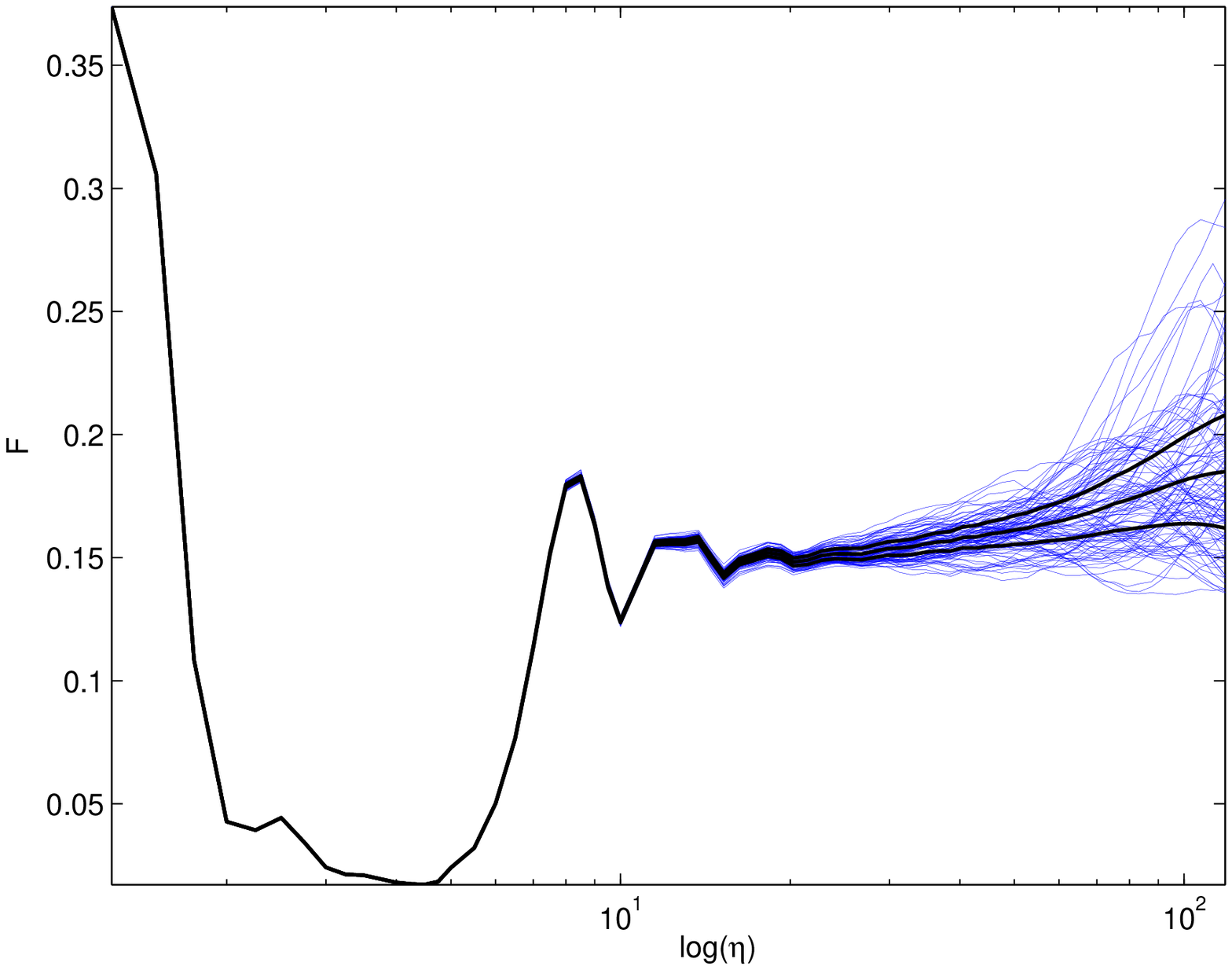}
\caption{\label{fig4} The same as Fig. \ref{fig1} but now for one hundred 
$256^3$ $3D$ radiation era simulations.}
\end{figure}

A total of several thousand matter and radiation era simulations in two and 
three spatial dimensions were run for various box sizes and 
dynamical ranges. Here, we compare some of these numerical results with 
analytic expectations and discuss the main results. In Fig. \ref{fig1} 
we plot the evolution of 
$R \equiv \eta A/V$ and $F$ for one hundred $2D$ matter era simulations. 
We also 
plot the average evolution of these parameters and the expected 1-sigma 
interval around the mean (using the analytic estimates for 
$\sigma_R$ and $\sigma_F$ described in the previous section). 
 Fig. \ref{fig2} shows 
a similar plot but now for one hundred $4096^2$ $2D$ radiation era 
simulations. Figs. \ref{fig3} and \ref{fig4} display analogous results 
for one hundred $256^3$ $3D$ matter and radiation era simulations 
respectivelly.

\begin{figure}
\includegraphics[width=2.8in]{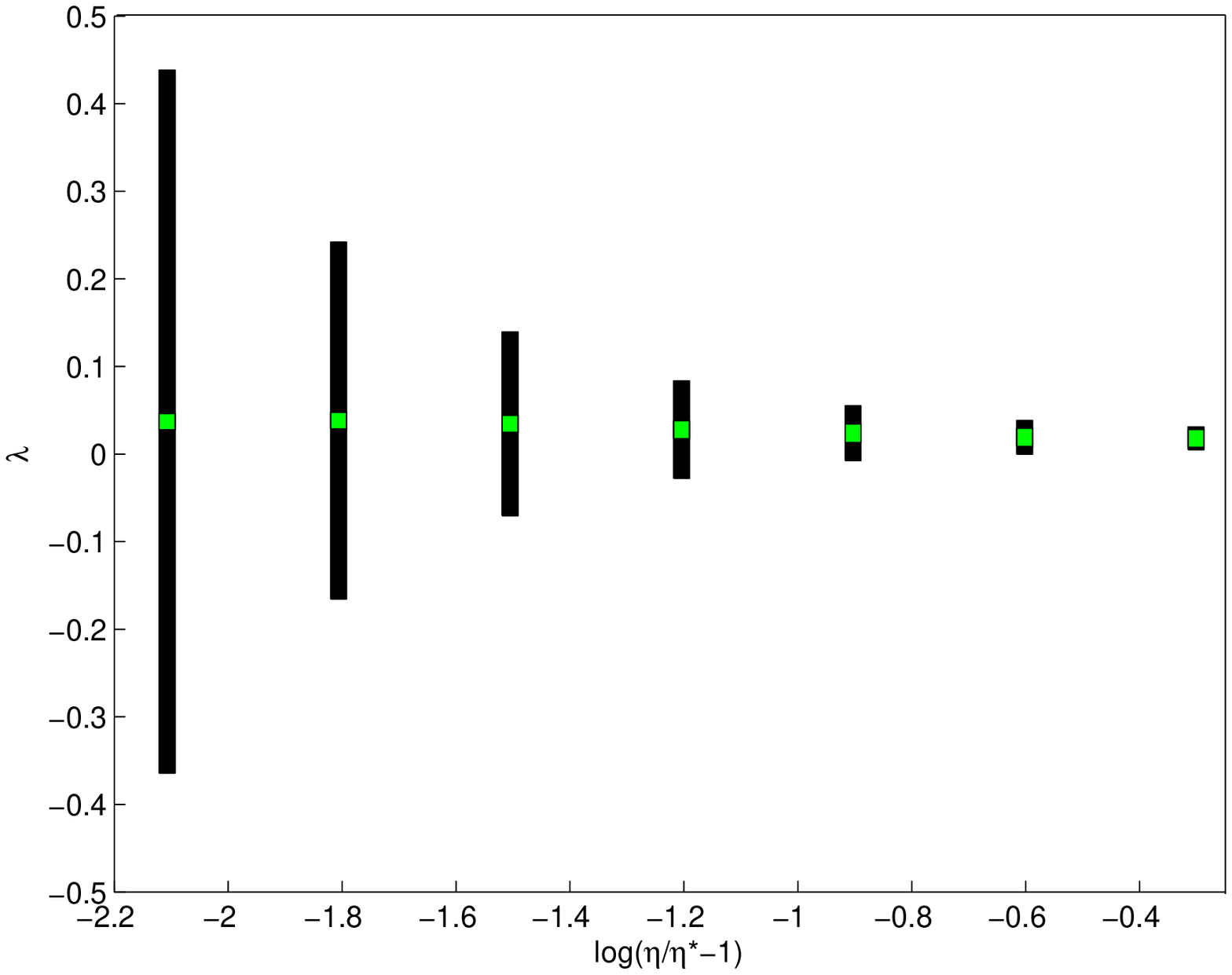}
\includegraphics[width=2.8in]{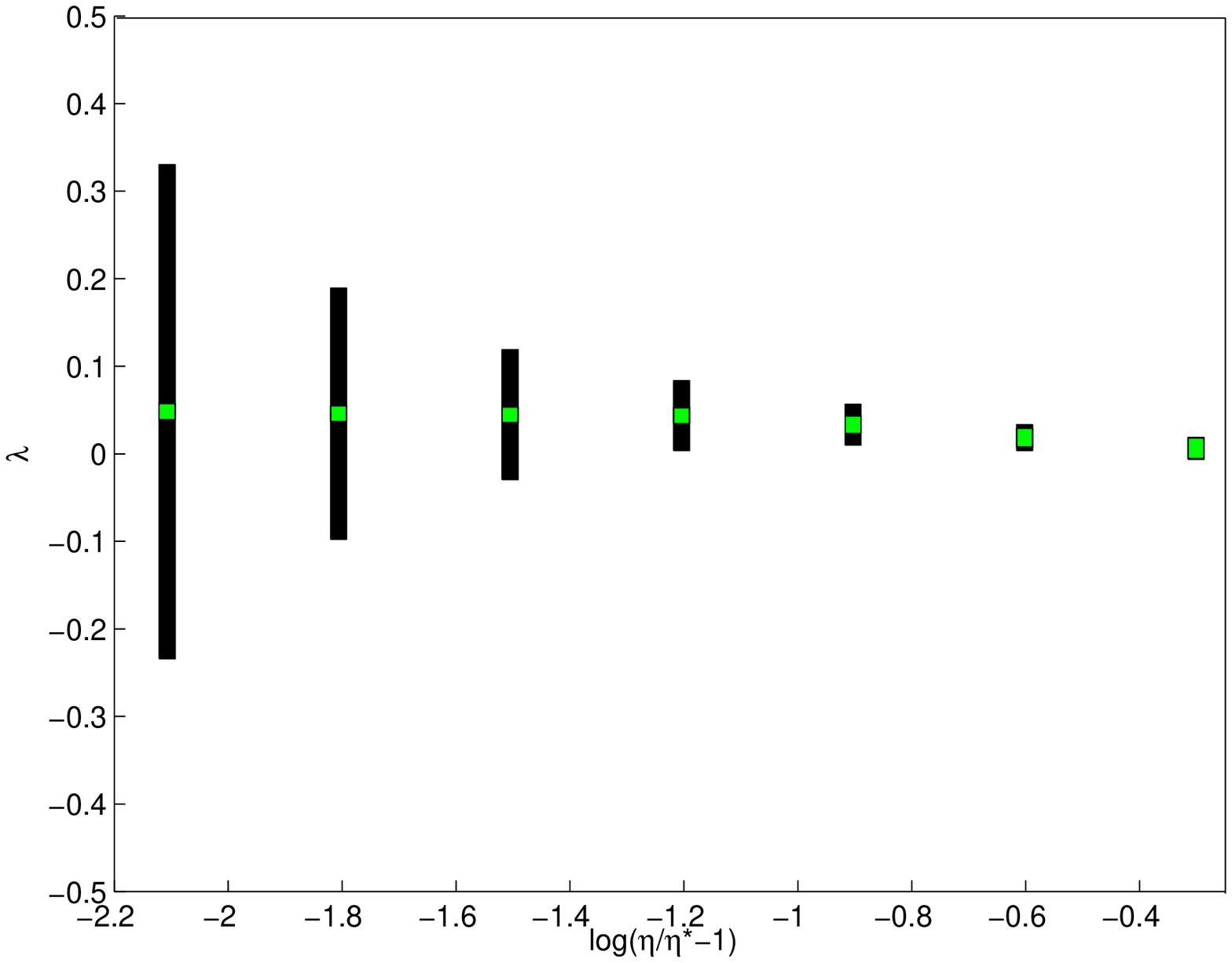}
\caption{\label{fig5} Comparison between numerical estimates of the 
standard deviation of the scaling parameter in an ensemble of $N_S$ 
simulations, $s_\lambda/{\sqrt {N_S}}$, with $s_\lambda$ computed either 
directly from the simulation (inner bars) or calculating $s_\Delta$ 
from the simulations and using eqn. (\ref{slambda}) to compute 
$s_\lambda$ assuming no correlation (outer bars) for 
$\eta_1=\eta_*=8^{-N_D/2}L$ and various values of $\eta_2=\eta$ for one 
hundred $4096^2$ $2D$ and one hundred $256^3$ $3D$ matter era 
simulations (left and right plots respectivelly).}
\end{figure}

\begin{figure}
\includegraphics[width=2.8in]{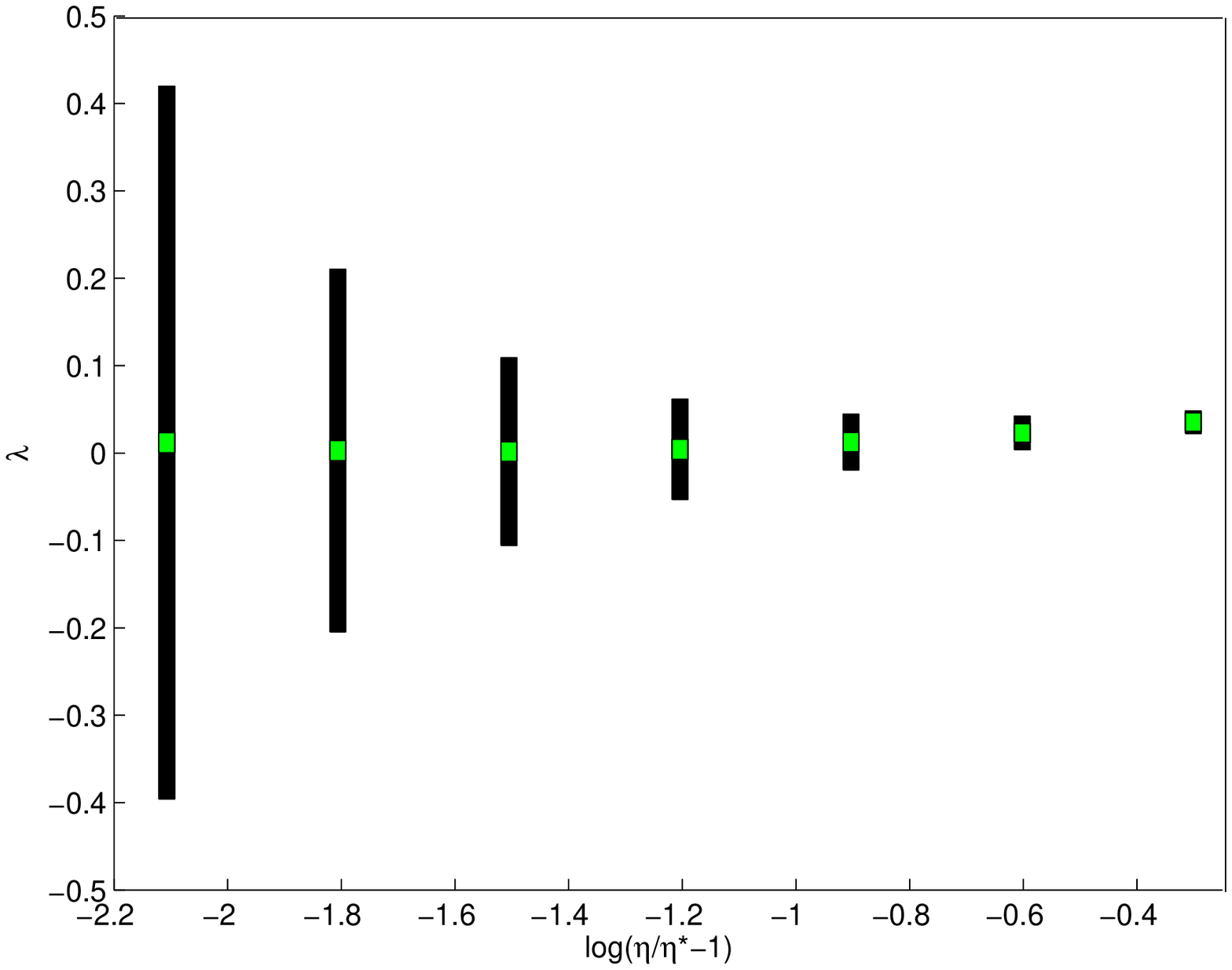}
\includegraphics[width=2.8in]{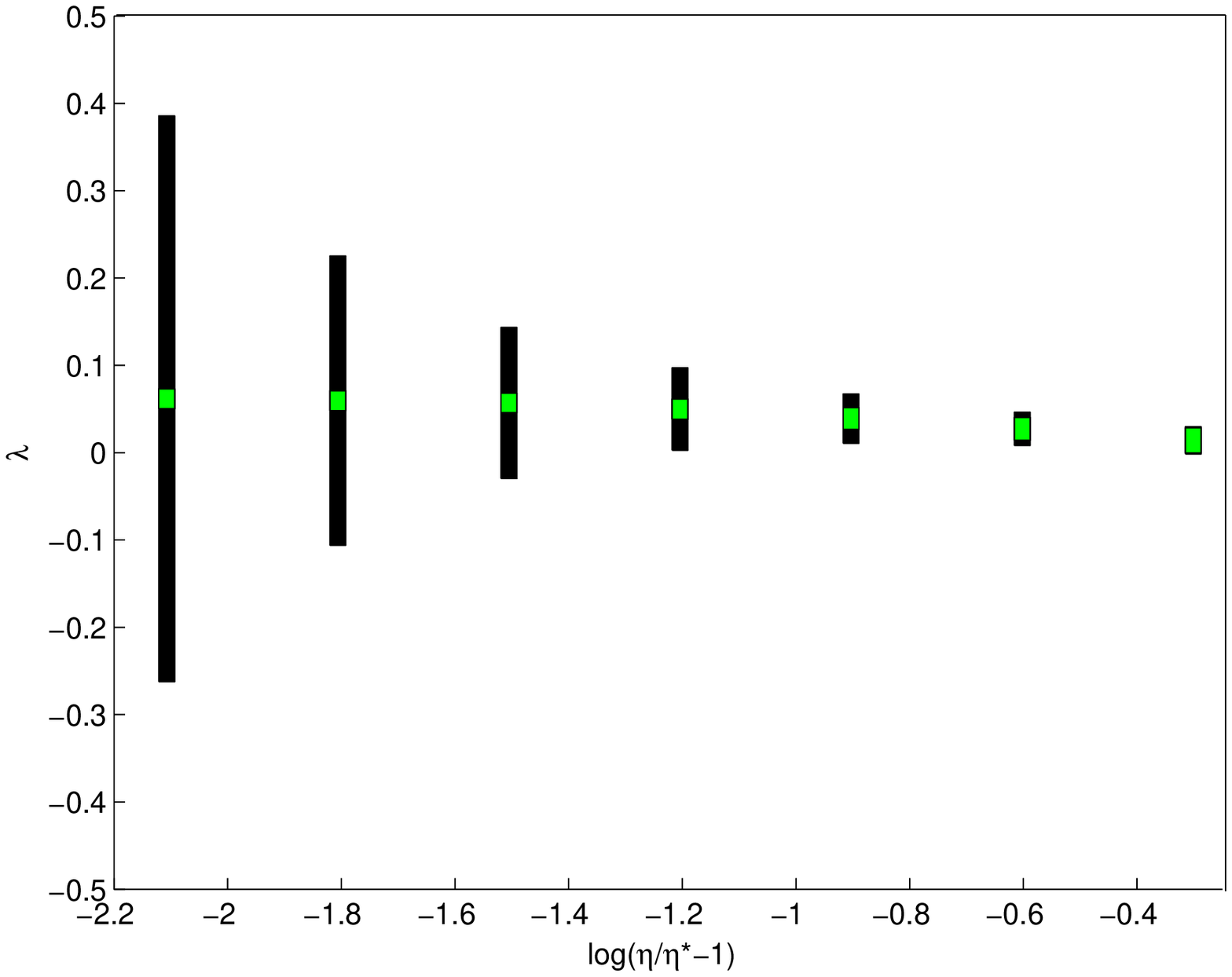}
\caption{\label{fig6}  The same a Fig. \ref{fig5} but now for one hundred 
$4096^2$ $2D$ and one hundred $256^3$ $3D$ radiation era 
simulations (left and right plots respectivelly). As expected 
when we increase $\eta_2-\eta_1$ our assumption of uncorrelation becomes 
increasingly more justified. The error bars are always placed in the 
middle of the interval $[\eta_1,\eta_2]$.}
\end{figure}

\begin{figure}
\includegraphics[width=2.8in]{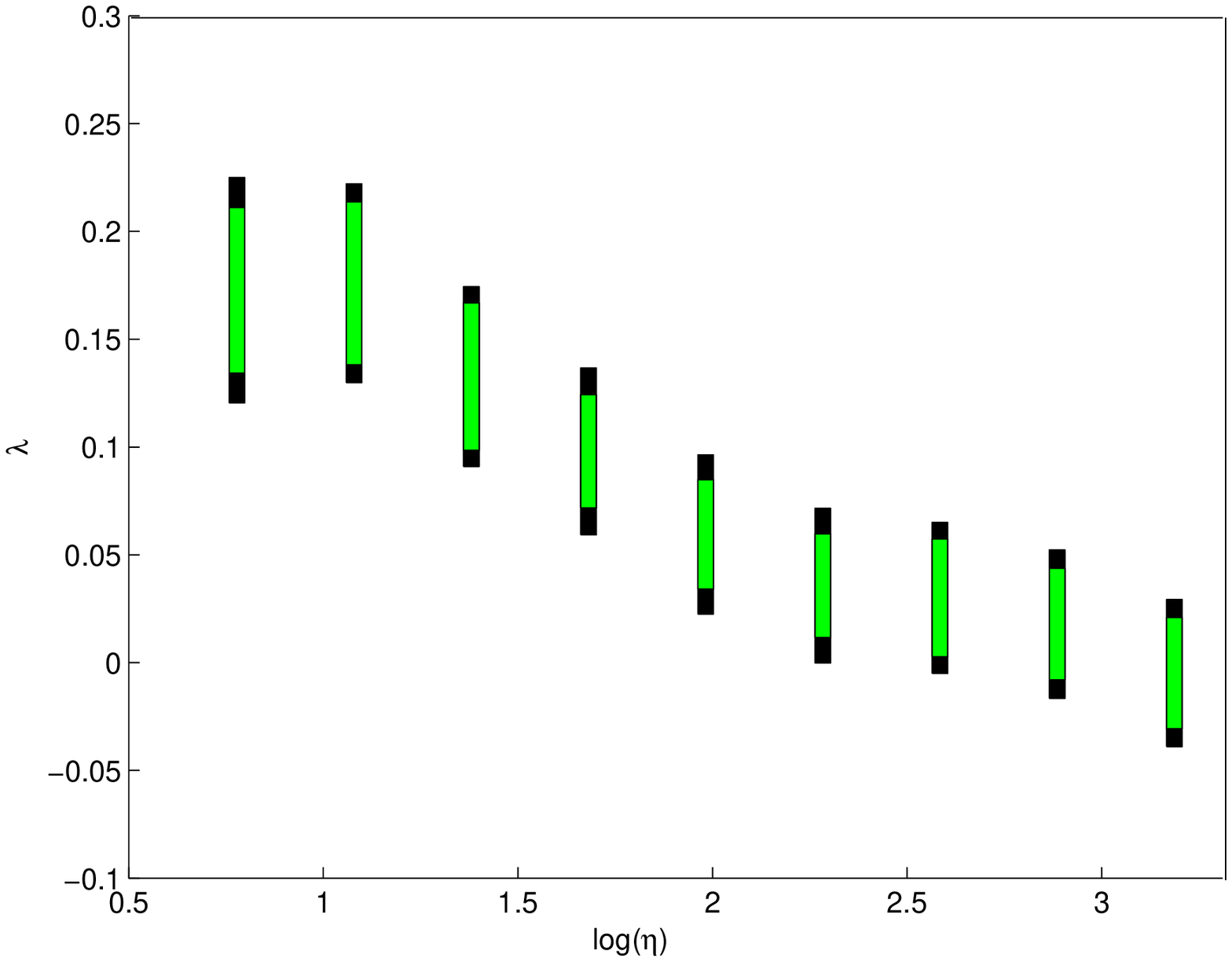}
\includegraphics[width=2.8in]{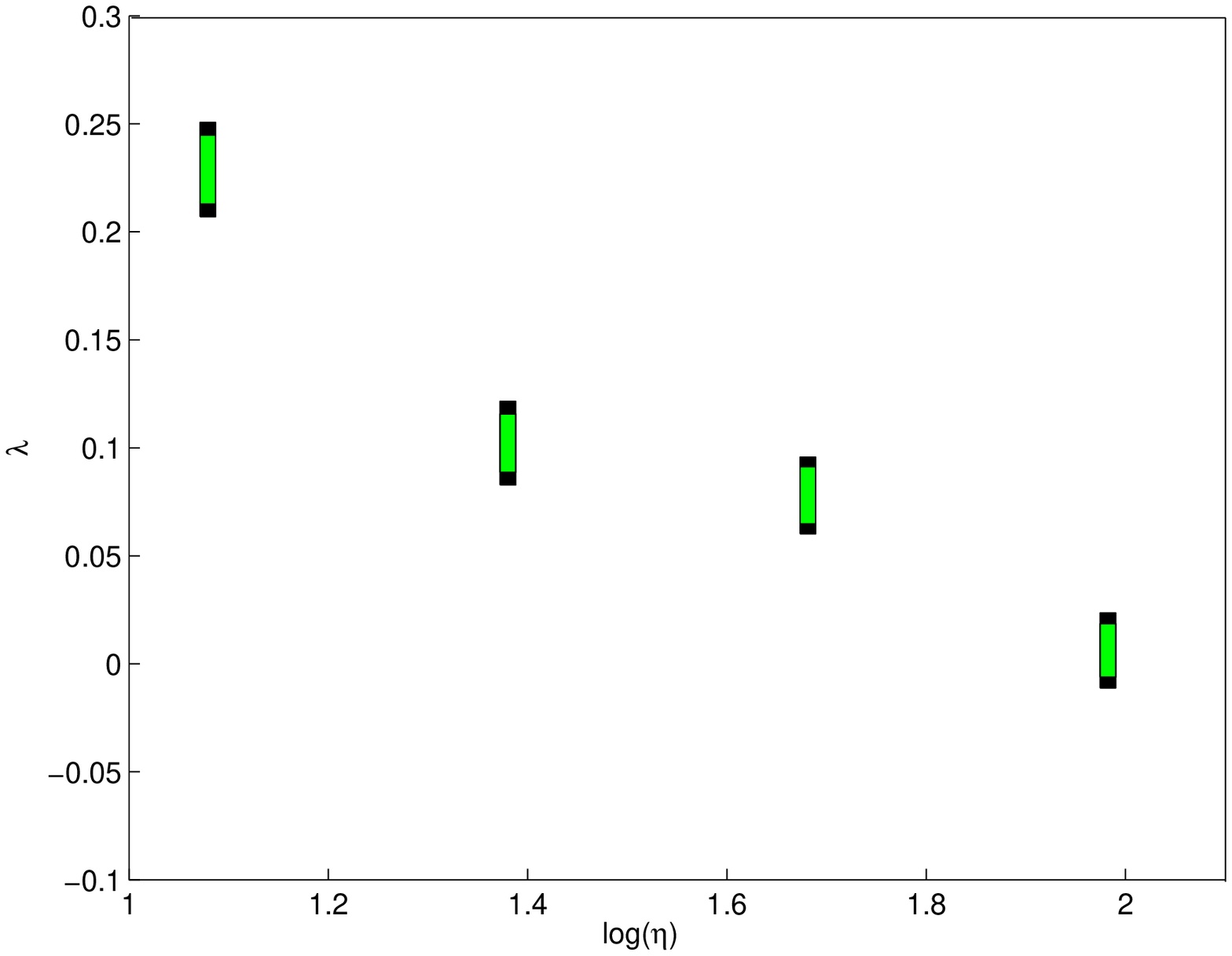}
\caption{\label{fig7} Evolution of the scaling exponent $\lambda$ 
as a function of the conformal time $\eta$ for one hundred $4096^2$ 
$2D$ and one hundred $256^3$ $3D$ matter era 
simulations (left and right plots respectivelly). The binnings have constant 
dynamical range (with $\eta_2/\eta_1=2$) and the error bars are 
the standard deviation of the scaling parameter in an ensemble of $N_S$ 
simulations, $s_\lambda/{\sqrt {N_S}}$, calculated either using our 
analytic estimate for $\sigma_\lambda$ (outer bars)
or by calculating $s_\Delta$ from the simulations 
and using eqn. (\ref{slambda}) to compute $s_\lambda$ assuming no correlation 
(inner bars). All the error bars (except the last one in each plot) 
were artificially enlarged by a factor of $(\eta/\eta_f)^{-N_D/2}$ where 
$\eta_f$ is the conformal time corresponding to the last error bar.
}
\end{figure}

\begin{figure}
\includegraphics[width=2.8in]{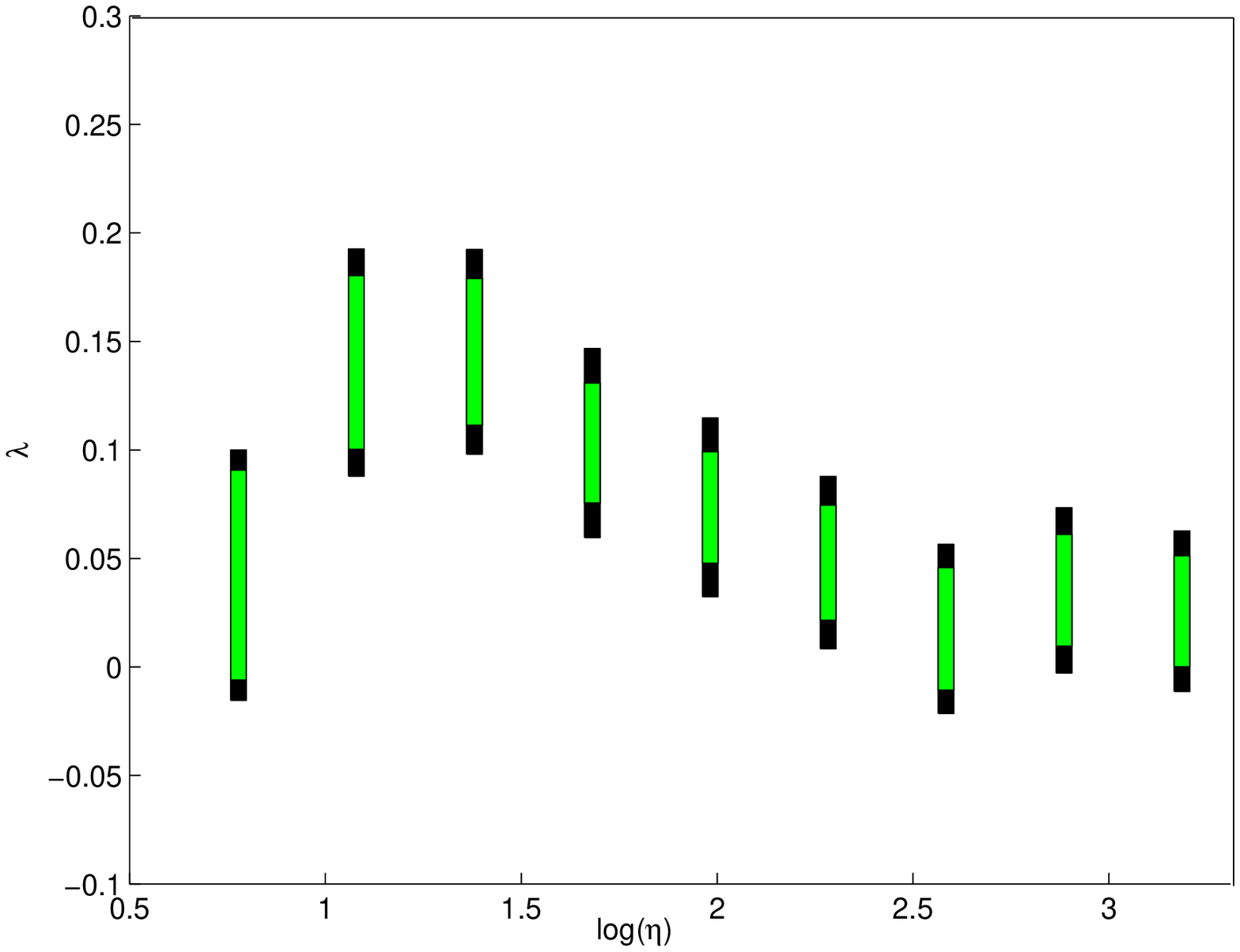}
\includegraphics[width=2.8in]{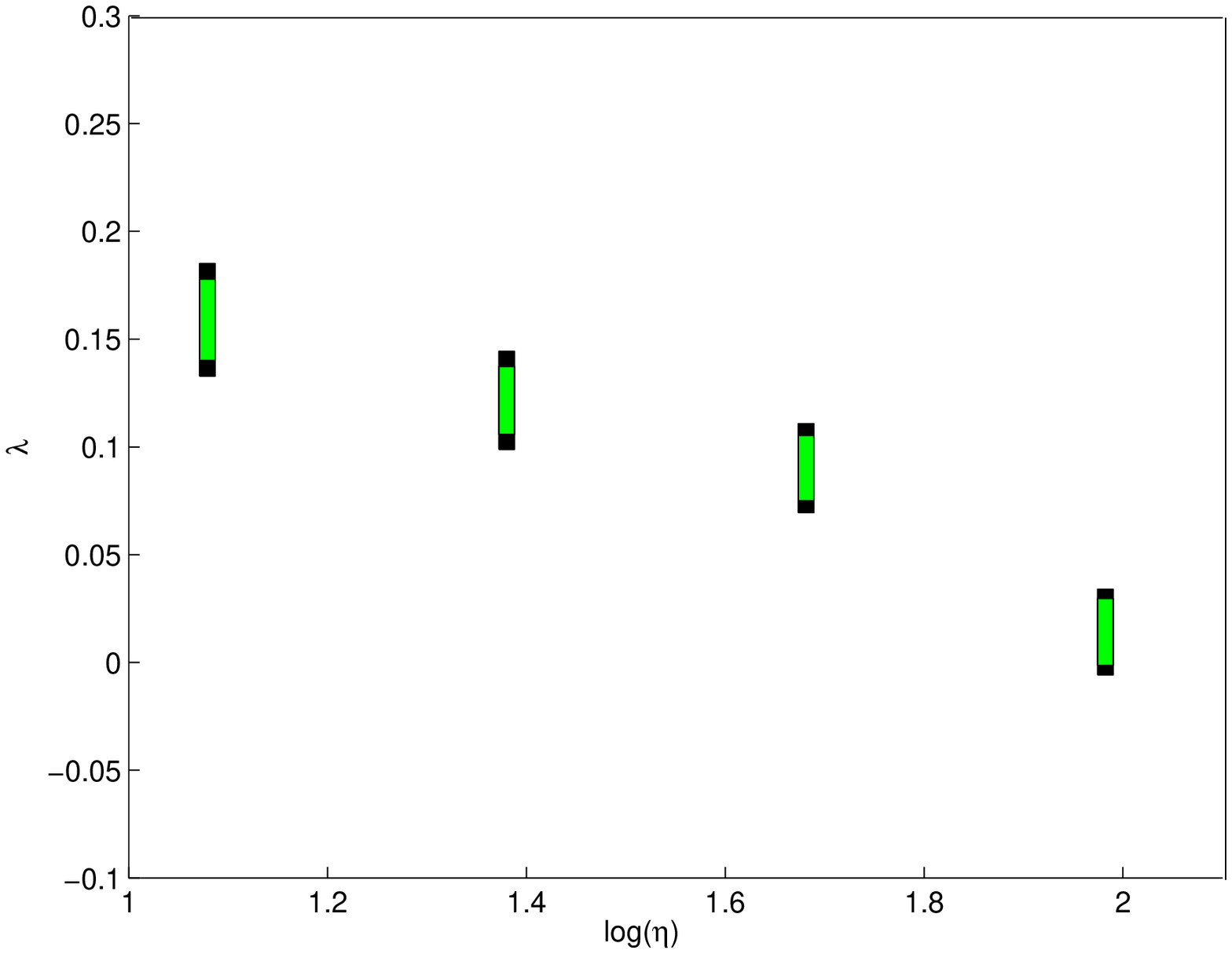}
\caption{\label{fig8} The same a Fig. \ref{fig7} but now for one hundred 
$4096^2$ $2D$ and one hundred $256^3$ $3D$  radiation era 
simulations (left and right plots respectivelly).}
\end{figure}

To zeroth order the results do not seem to be very sensitive on the number 
of spatial dimensions, $N_D$. 
We clearly see that although the network is not scaling during most of 
the simulation it seems to approach a scaling solution very slowly at late 
times.
Also, the main difference between matter and radiation era simulations lies in 
the larger value of the typical velocity of the domain walls in the latter 
case although we also find a considerable difference in the evolution of 
the matter and radiation era runs at early times when the initial conditions 
are still important.
 
As expected we see that in the case of the domain wall area the analytic 
estimate is a good approximation ($\sigma_R$ is only slightly overestimated - 
see Figs. \ref{fig7} and \ref{fig8}) while in the case of the parameter $F$ 
the uncertainty is underestimated due to not taking into account fact the 
uncertainty in the velocity of the domain walls as well as the part of the 
kinetic energy which is not associated with the domain walls themselves but 
with the particle radiation emitted by them.

Our analytic estimates for the variance of the scaling exponent, $\lambda$, 
were based on the assumption that the statistical properties of the 
domain wall network at two different times $\eta_1$ and $\eta_2$ were 
uncorrelated. Here we test for the validity of that assumption for different 
sizes of the conformal time interval $\eta_2-\eta_1$. 
In Fig. \ref{fig5}  we compare the numerical estimates of the standard 
deviation of the scaling parameter in an ensemble of $N_S$ 
simulations, $s_\lambda/{\sqrt {N_S}}$, with $s_\lambda$ computed either 
directly from the simulation (inner bars) or calculating $s_\Delta$ 
from the simulations and using eqn. (\ref{slambda}) to compute 
$s_\lambda$ assuming no correlation (outer bars) for 
$\eta_1=\eta_*=8^{-N_D/2}L$ and various 
values of $\eta_2=\eta$ for one hundred $4096^2$ $2D$ and one hundred $256^3$ 
$3D$ matter era simulations (left and right plots respectivelly). We 
clearly see that when 
we increase $\eta_2-\eta_1$ our assumption of no correlation becomes 
increasingly more justified and that there is no significant dependence of the 
correlation time on the number of spatial dimensions $N_D$. Fig. \ref{fig6} 
displays analogous results but now for radiation era simulations. However, no 
significant differences from the matter era runs are found.

In Fig. \ref{fig7} we plot the evolution of the scaling exponent $\lambda$ 
as a function of the conformal time $\eta$ for one hundred $4096^2$ 
$2D$ and one hundred $256^3$ $3D$ matter era simulations (left and right 
plots respectivelly). 
Here, we chose binnings of constant dynamical range (with $\eta_2/\eta_1=2$) 
and the error bars are 
the standard deviation of the scaling parameter in an ensemble of $N_S$ 
simulations, $s_\lambda/{\sqrt {N_S}}$, calculated 
either using our analytic estimate for $\sigma_\lambda$ (outer bars) 
or by calculating $s_\Delta$ from the simulations 
and using eqn (\ref{slambda}) to compute $s_\lambda$ assuming no correlation 
(inner bars). Note that the error bars in  Fig. \ref{fig7} were 
artificially enlarged  by a factor of of $(\eta/\eta_f)^{-N_D/2}$ where 
$\eta_f$ is the conformal time corresponding to the last error bar on the 
right so that at early times the real 
error bars are much smaller than at late times. We clearly see that our 
simple model successfully 
predicts the evolution of the uncertainties in the value of the 
scaling exponent, $\lambda$, although the performance of the method 
may slightly depend on the number of spatial dimensions, $N_D$. We find 
that, as expected, at early times there are 
strong deviations from the linear scaling solution but the networks seem 
to slowly approach a linear scaling solution at late times. 
Fig. \ref{fig8} displays analogous results but now 
for radiation era simulations but no significant 
differences (as far as the approach to scaling is concerned) 
were found.

Our results are consistent with those of previous studies 
\cite{Press,Coulson,Larsson,Fossils,Garagounis} in finding
a slow approach to linear scaling as well as strong deviations from a linear
scaling solution at early times. However, we do not find any evidence for
strong deviations from a linear scaling solution at late times. We note
that this is not in disagreement with the results obtained by other authors
but it is a consequence of the larger number, size and dynamical
range of the simulations analysed in our paper.

We have also performed a small number of $128^4$ $4D$ simulations which in 
a simple phenomenologogical way be of interest to brane world scenarios 
\cite{Oliveira,tye0,Brax}. The results obtained are consistent with those 
described in this paper for the $2D$ and $3D$ simulations.

This work was partially funded by Funda{\c c}\~ao para a Ci\^encia
e Tecnologia (Portugal) under contract POCTI/FP/FNU/50161/2003.
We thank Ruth Durrer and Paul Shellard for useful discussions and 
suggestions. J.O. is grateful for the hospitality of DAMTP (Cambridge), 
where some of the present work was carried out. This work was done in the 
context of the ESF COSLAB network, and was performed on COSMOS, the 
Altix3700 owned by the UK Computational Cosmology Consortium, 
supported by SGI, Intel, HEFCE and PPARC.

\end{document}